
\input epsf

\ifx\epsffile\undefined\message{(FIGURES WILL BE IGNORED)}
\def\insertfig#1#2{}
\else\message{(FIGURES WILL BE INCLUDED)}
\def\insertfig#1#2{{{\baselineskip=4pt
\midinsert\centerline{\epsfxsize=\hsize\epsffile{#2}}{{
\centerline{#1}}}\medskip\endinsert}}}
\fi
\input harvmac
%
%
%
%
\ifx\answ\bigans
\else
\output={

\almostshipout{\leftline{\vbox{\pagebody\makefootline}}}\advancepageno

}
\fi
%
%
%

%
%

%
%
\def\UCSD#1#2{\noindent#1\hfill #2%
\bigskip\supereject\global\hsize=\hsbody%
\footline={\hss\tenrm\folio\hss}}
%
%
\def\abstract#1{\centerline{\bf Abstract}\nobreak\medskip\nobreak\par
#1}
%
%
%
%
\edef\tfontsize{ scaled\magstep3}
 \tfontsize  \tfontsize
\font\titlermss=cmr5 \tfontsize \font\titlei=cmmi10 \tfontsize
\font\titleis=cmmi7 \tfontsize \font\titleiss=cmmi5 \tfontsize
\font\titlesy=cmsy10 \tfontsize \font\titlesys=cmsy7 \tfontsize
\font\titlesyss=cmsy5 \tfontsize  \tfontsize
\skewchar\titlei='177 \skewchar\titleis='177 \skewchar\titleiss='177
\skewchar\titlesy='60 \skewchar\titlesys='60 \skewchar\titlesyss='60
\scriptscriptfont0=\titlermss
\scriptscriptfont1=\titleiss
\scriptscriptfont2=\titlesyss
%
%
%

%
\def\inv{^{\raise.15ex\hbox{${\scriptscriptstyle -}$}\kern-.05em 1}}
\def\lbar{{\lower.35ex\hbox{$\mathchar'26$}\mkern-10mu\lambda}}

%
%
%
%
\def\dsl{\,\raise.15ex\hbox{/}\mkern-13.5mu D} 
\def\delsl{\raise.15ex\hbox{/}\kern-.57em\partial}
\def\Ksl{\hbox{/\kern-.6000em\rm K}}
\def\Asl{\hbox{/\kern-.6500em \rm A}}
\def\Dsl{\hbox{/\kern-.6000em\rm D}} 
\def\Qsl{\hbox{/\kern-.6000em\rm Q}}
\def\gradsl{\hbox{/\kern-.6500em$\nabla$}}
%
%
\def\lspace{\ifx\answ\bigans{}\else\qquad\fi}
\def\lbspace{\ifx\answ\bigans{}\else\hskip-.2in\fi} 
%
%
\def\boxeqn#1{\vcenter{\vbox{\hrule\hbox{\vrule\kern3pt\vbox{\kern3pt
        \hbox{${\displaystyle #1}$}\kern3pt}\kern3pt\vrule}\hrule}}}
%
%
\def\mbox#1#2{\vcenter{\hrule \hbox{\vrule height#2in
\kern#1in \vrule} \hrule}}
%
%
%
%

  \def\CO{{\cal O}}

%
%
%
%
%

%

\def\bar#1{\overline{#1}}

\def\darr#1{\raise1.5ex\hbox{$\leftrightarrow$}\mkern-16.5mu #1}

%
%
\def\frac#1#2{{\textstyle{#1\over #2}}} 
%
%
%
%

\def\MeV{{\rm MeV}}

%
%
%
%

%
%
\def\ltap{\ \raise.3ex\hbox{$<$\kern-.75em\lower1ex\hbox{$\sim$}}\ }
\def\gtap{\ \raise.3ex\hbox{$>$\kern-.75em\lower1ex\hbox{$\sim$}}\ }
\def\gl{\ \raise.5ex\hbox{$>$}\kern-.8em\lower.5ex\hbox{$<$}\ }
\def\roughly#1{\raise.3ex\hbox{$#1$\kern-.75em\lower1ex\hbox{$\sim$}}}

%
%

%

%
\def\np#1#2#3{{Nucl.\ Phys.} B{#1} (#2) #3}
\def\pl#1#2#3{{Phys.\ Lett.} {#1}B (#2) #3}
\def\prl#1#2#3{{Phys.\ Rev.\ Lett.} {#1} (#2) #3}
\def\physrev#1#2#3{{Phys.\ Rev. } {D#1} (#2) #3}

\relax

\def\hbar{\bar h_Q}

\def\qsl{\hbox{/\kern-.5600em {$q$}}}
\def\ksl{\hbox{/\kern-.5600em {$k$}}}

\def\({\left(}
\def\){\right)}

\def\OMIT#1{}
\def\frac#1#2{{#1\over#2}}

\def\blmscale{\mu_{\rm BLM}}
\def\OMIT#1{}
\def\alphams{\overline{\alpha}_s(m_b)}
\def\alphabar{\overline{\alpha}_s}
\def\msbar{$\overline{\rm MS}$\ }

\hbadness=10000

\noblackbox
\vskip 1.in
\centerline{{\titlefont{Charm Mass Dependence of the $\CO(\alpha_s^2 n_f)$}}}
\centerline{{\titlefont{Correction to Inclusive $B\rightarrow X_ce\bar\nu_e$
Decay}}}
\medskip
\vskip .5in
\centerline{Michael Luke${}^{a}$, Martin J.~Savage${}^{b}$
and Mark B.~Wise$^{c}$}
\medskip
{\it{
\centerline{a) Department of Physics, University of Toronto,
Toronto,
Canada M5S 1A7}
\centerline{b) Department of Physics, Carnegie Mellon
University,
Pittsburgh PA 15213}
\centerline{c) Department of Physics, California Institute
of Technology,
Pasadena, CA 91125}}}

\vskip .2in

\abstract{
We compute the $\alpha_s^2 n_f$ perturbative QCD contribution to semileptonic B
decay,
including the finite mass of the charm quark.  This result provides an estimate
of the
size of the two-loop correction, which is found to be about 50\% of the
one-loop
correction. We use these results to set the scale for the one-loop correction
using
the   scheme of Brodsky, Lepage and Mackenzie and find a BLM scale of $\mu_{\rm
BLM}
= 0.13\, m_b$, when the inclusive semileptonic rate is expressed in terms of
the $b$
and $c$ quark pole masses and the \msbar strong coupling.  The two loop
correction lies roughly midway between that obtained at
$m_c=0$ and that obtained in the Shifman-Voloshin limit $m_b,
m_c\gg m_b-m_c\gg\Lambda_{\rm QCD}$ while the corresponding BLM scale is
somewhat closer to that obtained in the former case.}
\vfill
\UCSD{\vbox{
\hbox{UTPT 94-27}
\hbox{CMU-HEP 94-32}
\hbox{CALT-68-1955}
\hbox{DOE-ER/40682-86}}
}{October 1994}
\eject

\newsec{Introduction}

Inclusive semileptonic $B$ decays provide a method  for
determining the magnitude of the element of the Cabibbo--Kobayashi--Maskawa
matrix $V_{cb}$.
In the limit where the $b$ quark mass is much larger than the QCD scale the $B$
meson decay rate is equal to the $b$ quark decay rate \ref\cgg{J.~Chay,
H.~Georgi and
B.~Grinstein, \pl{247}{1990}{399}.}. Corrections to this
first arise at order
$(\Lambda_{QCD}/m_b)^2$ and these nonperturbative corrections may be written in
terms of the matrix elements
\ref\bsuv{M.\ Voloshin and M.\ Shifman, Sov.\ J.\ Nucl.\ Phys.\ 41 (1985) 120;
I.\ I.\ Bigi, N.\ G.\ Uraltsev and A.\ I.\ Vainshtein, \pl{293}{1992}{430};
B.\ Blok, L.\ Koyrakh, M.\ Shifman and A.\ I.\ Vainshtein, \physrev{49}
{1994}{3356};  I.\ I.\ Bigi, M.\ Shifman, N.\ G.\ Uraltsev and A.\ I.\
Vainshtein, \prl
{71}{1993}{496}.}%
\nref\mb{A.\ V.\ Manohar and M.\ B.\ Wise, \physrev{49} {1994}{1310}.}%
--\ref\man{T. Mannel, \np{413}{1994}{396}.}
$\langle B| \bar b (iD)^2 b|B\rangle$ and
$\langle B| \bar b ig \sigma_{\mu\nu} G^{\mu\nu} b|B\rangle$.
The order $\alpha_s$ corrections to the decay rate, including the complete $c$
quark mass dependence, have been calculated in
\ref\jk{M. Jezabek and J.H. Kuhn, \np{314}{1989}{1}.}.

No complete calculation of the two-loop perturbative corrections to inclusive
heavy quark decay has been performed.  However, the $\CO(\alpha_s^2 n_f)$
correction to the inclusive semileptonic decay rate to a massless quark has
been
recently calculated
\ref\mmm{M.\ Luke, M.\ J.\ Savage and M.\ B.\ Wise, University of Toronto
Preprint UTPT 94-24 (1994).}.  This correction is of interest in its own right
for two
reasons.  First, if one adopts the viewpoint of Brodsky, Lepage and Mackenzie
\ref\blm{S.\ J.\ Brodsky, G.\ P.\ Lepage and P.\ B.\ MacKenzie,
\physrev{D28}{1983}{228}.}, one can use this result to set the appropriate
scale for
$\alpha_s$ in the leading term.  For $b\rightarrow X_u$ semileptonic decay,
this
leads to the surprisingly low scale $\blmscale=0.07\,m_b$ \mmm.
Second,
even if one does not adopt this viewpoint, we argue that this calculation
provides a
good estimate of the size of the two-loop calculation.  This is not because
$n_f$ is
large, but rather because the QCD beta function $\beta=11-2n_f/3$ is large, so
the
vacuum polarization graphs which contribute to this term are be expected to
dominate the two-loop result.  Empirically, this is
certainly true for
$R(e^+e^-\rightarrow {\rm hadrons})$,
$\Gamma(\tau\rightarrow \nu_\tau+{\rm hadrons})$ and the
two-loop relation between the pole mass and the \msbar mass of a heavy
quark:
\eqn\otherones{\eqalign{
&R(e^+e^-\rightarrow {\rm hadrons})=3\left(\sum_i Q_i^2\right)\left[
1+{\alphabar(\sqrt{s})\over\pi}+\left(0.17\beta+0.08\right)
\left({\alphabar(\sqrt{s})\over\pi}\right)^2+
...\right]\cr
&{\Gamma(\tau\rightarrow\nu_\tau+{\rm hadrons})\over
3\Gamma(\tau\rightarrow\nu_\tau\bar\nu_e e^-)}=1+{\alphabar(m_\tau)
\over\pi}+\left(0.57\beta+0.08\right)\left({\alphabar(m_\tau)\over\pi}\right)^2+
...
\cr
&{m_Q^{\rm pole}\over m_Q^{\overline {MS}}(m_Q)}=1+{4\over
3}{\alphabar(m_Q)\over\pi}+
\left(1.56\beta-1.05\right)\left(
{\alphabar(m_Q)\over\pi}\right)^2+ ...}}
where $\alphabar$ is the \msbar strong coupling.
In each of these cases, the $\CO(\alpha_s^2\beta)$ term provides an
excellent approximation to the full two-loop correction.  Using the term
proportional to
$\beta$ as an estimate of the size of the two-loop QCD corrections for
$b\rightarrow X_u$ semileptonic decay and taking
$n_f=3$\footnote{$^\dagger$}{We
have not taken into account the $c$ quark mass for virtual $c$ quarks.  The
true
result will be somewhere between $n_f=3$ and $n_f=4$, and we have chosen to
take
the number of light flavours to be three for both $c$ and $b$ decays.} gives
\mmm
\eqn\twoloopa{\eqalign{\Gamma (b\rightarrow X_ue\overline{\nu_e})  &=
|V_{bu}|^2
{G_F^2 m_b^5\over 192\pi^3}
\left[ 1 -\left({\alphams\over\pi}\right)[2.41] -
\left({\alphams\over\pi}\right)^2 [28.7]+...\right]\cr
&=|V_{bu}|^2 {G_F^2 m_b^5\over 192\pi^3}
[ 1-0.15-0.11+ ...]}}
where we have used $\alphams\simeq 0.2$.  The two-loop term is clearly
significant.

For charmed hadrons the situation is even worse.  Using $\alphabar(m_c)=0.29$,
we find
\eqn\twoloopb{\Gamma (c\rightarrow X_de\overline{\nu_e}) =|V_{cd}|^2 {G_F^2
m_c^5\over 192\pi^3}
[ 1-0.22-0.25+ ...]}
and it does not appear that the perturbation series is controlled at all. This
is
reflected in the small size of the corresponding BLM scale $\blmscale\sim
100\,\MeV$.

The results of \mmm\ raise
questions about the applicability of perturbative QCD to describe the
semileptonic
decay of charmed hadrons, and also suggest that two-loop corrections are
important for a reliable extraction of $V_{bc}$ from inclusive $b\rightarrow
X_c$
decays.   In
this paper we extend these results to include the effects of the finite $c$
quark
mass in $b\rightarrow X_c$ decays.  We find that, for
$m_c/m_b =0.3$, the
BLM scale is raised to $\blmscale=0.13\,m_b$, and the corresponding
perturbation
series is
\eqn\twoloopc{\eqalign{\Gamma (b\rightarrow X_ce\overline{\nu_e}) & =
|V_{bc}|^2
{G_F^2 m_b^5\over 192\pi^3} [0.52]
\left[ 1 \ -\  \left({\alphams\over\pi}\right)[1.67]-
\left({\alphams\over\pi}\right)^2 [15.1] + ...\right]\cr &=|V_{bc}|^2 {G_F^2
m_b^5\over 192\pi^3}
[ 1-0.11-0.06+ ...]}}
The finite charm quark mass thus  reduces the size of
the higher-order corrections, although they are still significant.

In Eqs.\ \twoloopa--\twoloopc, $m_b$ and $m_c$ are the pole masses of the $b$
and
$c$ quark.  If \msbar masses are used in \twoloopa\ and \twoloopb\ the
convergence is slightly improved \mmm.

\newsec{Calculation}

The semileptonic decay rate of a $b$-quark is given by
\eqn\semia{\eqalign{
\Gamma (b\rightarrow X_ce\overline{\nu_e}) =
|V_{bc}|^2 \,{G_F^2m_b^5\over 192\pi^3} f\left({m_c\over m_b}\right) &
\left[ 1 - \left( {\alphams\over\pi}\right) {2\over
3}\left(\pi^2-{25\over 4}
+ \delta_1\left({m_c\over m_b}\right)\right)
\right. \cr
& \left. - \left( {\alphams\over \pi}\right)^2\ \left(  \beta\,
\chi_\beta
\left({m_c\over m_b}\right) +\chi_0 \left({m_c\over m_b}\right) \right)
+\ ...\right]}}
where $\beta=11-{2\over 3}n_f$, $\overline{\alpha_s}$
is the $\overline{MS}$ strong coupling, $f(x) = (1-x^4)(1-8x^2+x^4) - 24x^4
\log(x)$ and $m_{b,c}$ are the heavy quark pole masses.
The function $\delta_1(x)$ has been
computed
in \jk, and  $\delta_1(0.3) = -1.11$.
In this work we compute the function $\chi_\beta (x)$.

The second order corrections to the semileptonic decay rate have been shown
\ref\sv{B.H. Smith and M.B. Voloshin, TPT-MINN-94/16-T (1994).}
to be calculable from the first order corrections when a finite gluon mass is
included.
If the first order correction with a finite gluon mass $m_g$ is denoted by
$\Gamma^{(1)}(m_g)$
then the second order correction proportional to the perturbative
QCD beta function is
\eqn\twoa{
\Gamma^{(2)}_\beta = -\beta {\alpha_s^{(V)}(m_Q)\over 4\pi}
\int_0^\infty\ {d\mu^2\over\mu^2}\
\left( \Gamma^{(1)}(\mu) - {m_Q^2\over\mu^2+m_Q^2}  \Gamma^{(1)}(0)\right)
\ \ \ ,}
where $\alpha_s^{(V)}$ is the strong coupling constant evaluated in the
``V-scheme"
of \blm\ and is related to the \msbar coupling constant
$\alphabar$ by
\eqn\msbar{
\alpha^{(V)}_s(\mu) = \overline{\alpha}_s(\mu) +
{5\over 3}{\overline{\alpha}_s(\mu)^2\over 4\pi}\beta
+...\ .}

Both virtual and bremstrahlung graphs contribute to the
first order correction.  The virtual graphs were computed analytically, with
the
final integrations of the lepton $q^2$ and $m_{\rm gluon}$ performed
numerically, while
the phase space integrals for the bremstrahlung graphs were performed
numerically.

The leading two-loop correction $\chi_\beta (m_c/m_b)$ is
plotted in
\fig\chigraph{The function $\chi_\beta(m_c/m_b)$ from Eq.~\semia\ plotted vs.
$m_c/m_b$.}. According to the BLM scheme \blm, this term is absorbed into the
first
order correction by redefining the scale at which the first order result is
evaluated
to $\mu=\blmscale$, where
\eqn\blmsc{
\blmscale = m_b\exp\left[ -{3\chi_\beta (m_c/m_b)
\over \pi^2-{25\over 4} + \delta_1(m_c/m_b)}\right].}
The BLM scale is plotted as a function of $m_c/m_b$ in
\fig\scaleplot{The BLM scale as a function of $m_c/m_b$.}.
At $m_c=0$ we reproduce the result \mmm\ $\mu_{\rm BLM}(0)=0.07\,m_b$, while
for
$m_c/m_b=0.3$ we find
$\blmscale = 0.13\, m_b$.  This is somewhat higher than the
scale found for the case of massless final state quarks but is still
considerably smaller than $m_b$.

A non-trivial check on our numerical results is obtained in the
Shifman-Voloshin limit
$m_b, m_c\gg m_b-m_c\gg\Lambda_{\rm QCD}$ \ref\shvo{M.~Voloshin and
M.~A.~Shifman, Sov.\ Journ.\ Nuc.\ Phys.\ 47 (1988) 511.}. In this limit the
final hadronic
states are dominated by the $D$ and $D^*$ mesons (which are degenerate to
leading order in
$1/m_c$).  They are produced almost at rest, and the rates for these {\it
exclusive}
processes can be calculated in the heavy quark effective theory with no free
parameters.  This gives
\eqn\exclusive{
\Gamma (B\rightarrow D,D^*e\overline{\nu_e}) = {G_F^2 (m_b-m_c)^5\over 60\pi^3}
\left[
|\eta_V|^2 + 3 |\eta_A|^2\right]+\CO\left({(m_b-m_c)^6\over m_b}\right)}
where the coefficients $\eta_A$ and $\eta_V$ arise in matching from the QCD
currents to the heavy quark currents
\eqn\curr{\eqalign{
\overline{c}\gamma^\mu b\rightarrow \eta_V\overline{h}_c\gamma^\mu h_b\
+\ ...\cr
\overline{c}\gamma^\mu\gamma_5 b\rightarrow \eta_A\overline{h}_c
\gamma^\mu\gamma_5 h_b
\ +\ ...\ ,}}
and in this limit the heavy $c$ and $b$ quarks have the same 4-velocity.
The matching coefficients have the perturbative expansion
\eqn\etapert{\eqalign{\eta_V&=1+{1\over
3}{\alphams\over\pi}\phi(m_c/m_b)+\left(
{\alphams\over\pi}\right)^2\left[{1\over 72}\phi(m_c/m_b)\beta
+...\right]+ ...\cr
\eta_A&=1+{1\over3}{\alphams\over\pi}\left[\phi(m_c/m_b)-2\right]+\left(
{\alphams\over \pi}\right)^2\left[\left({5\over
72}\phi(m_c/m_b)-{14\over 72}\right)\beta+...\right]+...\, .}}
where
the leading term was calculated in Refs.\  \shvo\  and
\ref\etaleading{ J.\ E.\ Paschalis and G.\ J.\ Gounaris,
\np{222}{1983}{473}\semi F.\ E.\ Close, G.\ J.\ Gounaris and J.\ E.\ Paschalis,
\pl{149}{1984}{209}} and the
$\CO(\alpha_s^2 \beta)$ terms were calculated in
Ref.\ \ref\neubertcurr{M. Neubert, CERN-TH.7454/94 (1994).}.
The function $\phi(z)$ is defined by
\eqn\phidef{\phi(z)=-3{1+z\over 1-z}\log z-6}
and so $\phi(1)=0$.  Furthermore, at $m_c=m_b$ the vector current is not
renormalized, and Eq.\ \etapert\ simplies to
\eqn\simpetapert{\eqalign{\eta_V&=1\cr
\eta_A&=1-{2\over 3}{\alphams\over\pi}-{7\over 36}\left(
{\alphams\over\pi}\right)^2\left[\beta+...\right]+...\,}}
which gives
\eqn\exclpert{\eqalign{\Gamma(B\rightarrow D,D^*e\bar\nu_e)=&
{G_F^2(m_b-m_c)^5\over 15\pi^3}\left(1-{\alphams\over\pi}-{7\over
24}
\left( {\alphams\over\pi}\right)^2\left[\beta+...\right]+...\right)\cr
&+\CO\left[{(m_b-m_c)^6\over m_b}\right].}}
The one loop term is in agreement with the results of Jezabek and Kuhn
\jk, while the two-loop term is in
agreement with the limiting value of our inclusive calculation.  This gives
a BLM scale in the SV limit of $\blmscale=0.56\,m_b$.

\newsec{Discussion and Conclusions}

The BLM scale is one measure of the size of higher order corrections.  However,
as we argued in the introduction, we can also use our results as an estimate of
the size of the two-loop corrections with $\mu=m_b$.
For $m_c/m_b=0.3$, the inclusive rate at leading order in $1/m_b$ is
\eqn\numbersa{\eqalign{
\Gamma (b\rightarrow X_ce\overline{\nu_e}) & = |V_{bc}|^2 {G_F^2 m_b^5\over
192\pi^3} [0.52]
\left[1 -\left({\alphams\over\pi}\right)[1.67]-
\left({\alphams\over\pi}\right)^2
[15.1]\ +\ ...\right]\cr
& = |V_{bc}|^2 {G_F^2 m_b^5\over 192\pi^3}[0.52]\left[1-0.11-0.06+...\right]\cr
& = |V_{bc}|^2 {G_F^2 m_b^5\over 192\pi^3}[0.52]\left[0.83+...\right]}}
where the factor of $0.52$ arises from the lowest order phase space factor
$f(0.3)$ and
again we have taken $\alphams=0.20$.
The one and two loop corrections  relative to the leading term are plotted
as functions of $m_c/m_b$ in \fig\magb{The $\CO(\alpha_s)$ (dashed line) and
$\CO(\alpha_s^2\beta)$ (solid line) corrections to the expansion of
$\Gamma(b\rightarrow X_c e\bar\nu_e$) relative to the leading term, plotted as
functions of
$m_c/m_b$.  We are using
$\alphams=0.20$ and $n_f=3$.}.  We can compare this with our previous results
for
$b\rightarrow X_u$ semileptonic decay, Eq.\ \twoloopa.
The finite charm quark mass somewhat
reduces the size of the two-loop term, but it is still significant, about 50\%
of the
$\CO(\alpha_s)$ correction.
The two loop correction lies roughly midway between that obtained at
$m_c=0$ and that obtained in the Shifman-Voloshin limit while the corresponding
BLM scale is somewhat closer to that obtained in the former case.

We have expressed the $b\rightarrow X_c$ decay rate in terms of the $b$ and $c$
quark
pole masses.  This is convenient since the difference in these masses is
determined by the
meson masses (up to corrections of $\CO(1/m_{c,b}))$:
\eqn\masses{m_b-m_c=m_B-m_D+\CO\left(1/m_{b,c}\right)}
and the function $f(m_c/m_b)$ is less sensitive to the sum of the quark masses
than the
difference.

For charm decays, including the finite mass of the
$s$ quark has a negligible effect on the magnitude of the two-loop corrections.
Because the $s$ quark is so light its pole mass is not a useful quantity.
However, for
illustrative purposes we have plotted the size of the one and two loop
corrections relative to
the leading term as functions of
$m_s/m_c$ in \fig\magc{The
$\CO(\alpha_s)$ (dashed line) and
$\CO(\alpha_s^2\beta)$ (solid line) corrections to the expansion of
$\Gamma(c\rightarrow X_s e\bar\nu_e$) relative to the leading term, plotted as
functions of $m_s/m_c$.  We are using
$\alphabar(m_c)=0.29$ and $n_f=3$.}, using $\alphabar(m_c)=0.29$.  Note that
for
small
$m_s$, the two loop correction is larger than the one loop correction.  For
$m_s/m_c=0.13$ the inclusive semileptonic decay rate is
\eqn\numbersc{\eqalign{
\Gamma (c\rightarrow X_se\overline{\nu_e}) & = |V_{cs}|^2 {G_F^2 m_c^5\over
192\pi^3} [0.87]
\left[1 -\left({\alpha_s\over\pi}\right)[2.08]-
\left({\alpha_s\over\pi}\right)^2 [22.7]+...\right]\cr  & = |V_{cs}|^2 {G_F^2
m_c^5\over
192\pi^3}[0.87]\left[1-0.19-0.19+...\right].}} The perturbation series is still
clearly
uncontrolled.

\OMIT{In conclusion, we have computed the part of the second order perturbative
QCD
corrections of the form $\alpha_s^2n_f$ including the finite charm quark mass.
We find that the charm mass somewhat reduces the relative size of these higher
order corrections
while still giving rise to a low scale at which to evaluate the first order
coupling
constant.
Despite the small scale the perturbative corrections to the semileptonic
$b$-decay
rate seem to be well behaved and converging.   A complete second order
calculation
is still required in order to have complete confidence in the pQCD
corrections.}

\bigskip

This research was supported in part by
the Department of Energy under contract DE--FG02--91ER40682 and
DE-FG03-92-ER40701 and by the Natural Sciences and Engineering Research
Council of Canada.

\listrefs
\listfigs
\vfill\eject
\insertfig{Figure 1}{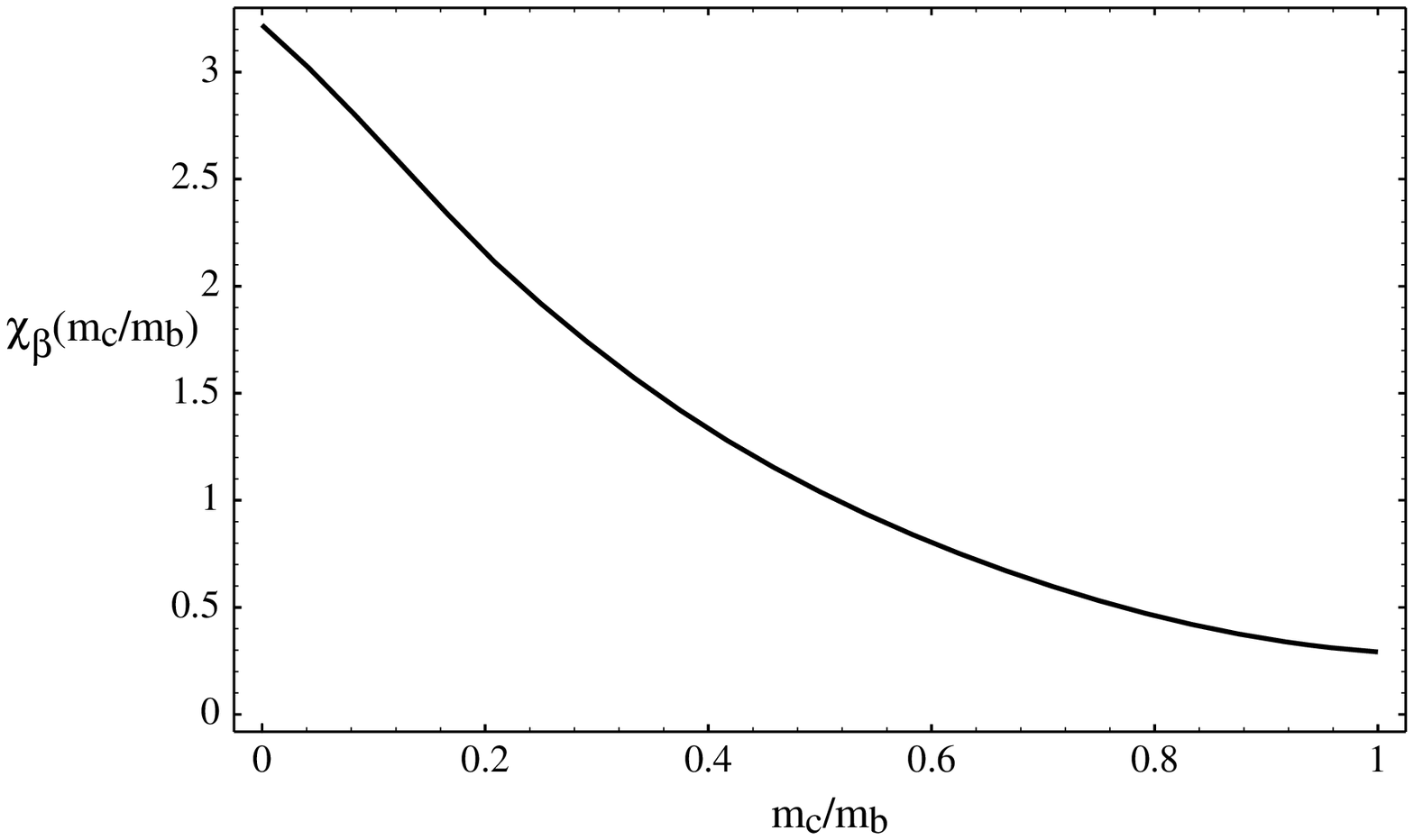}
\insertfig{Figure 2}{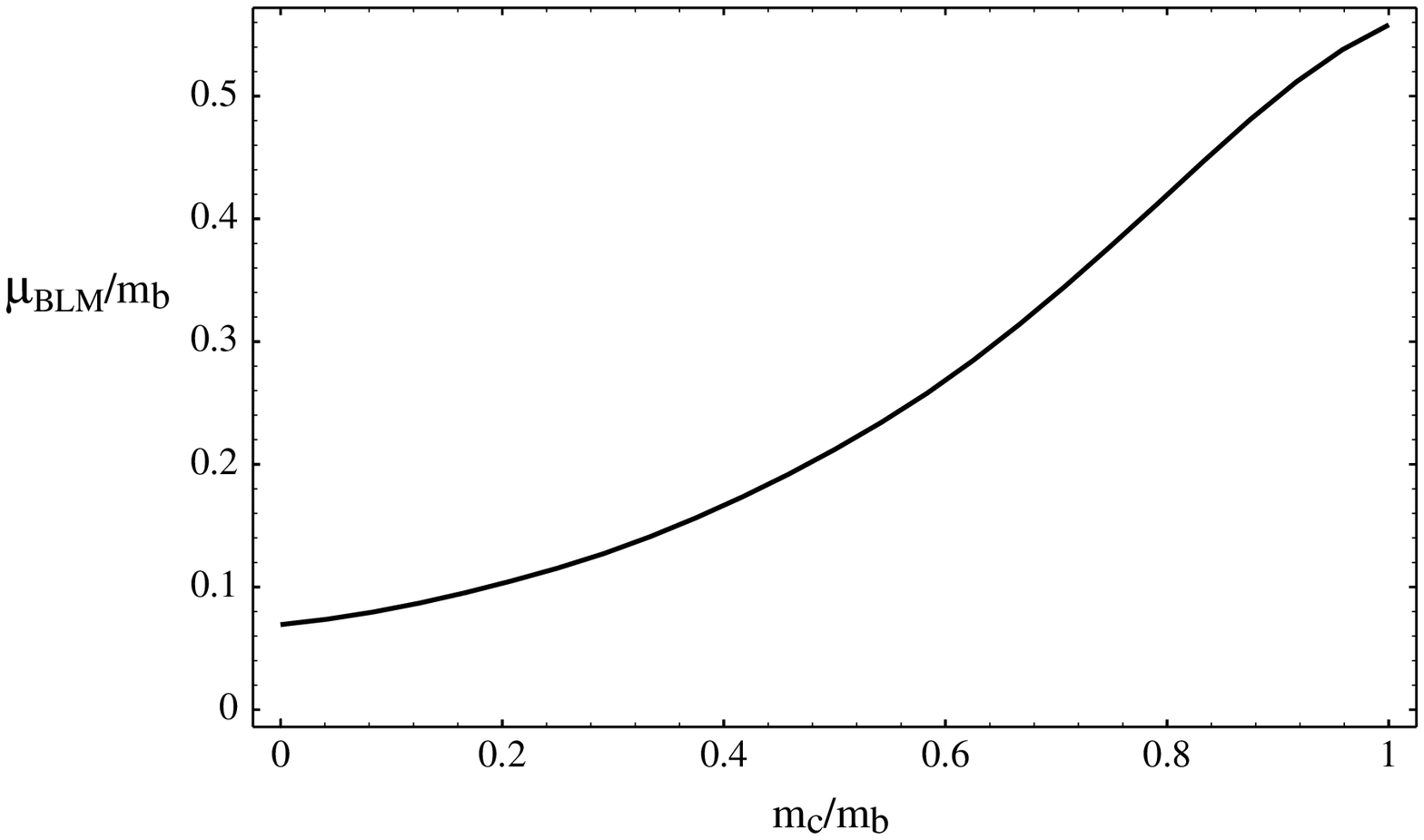}
\insertfig{Figure 3}{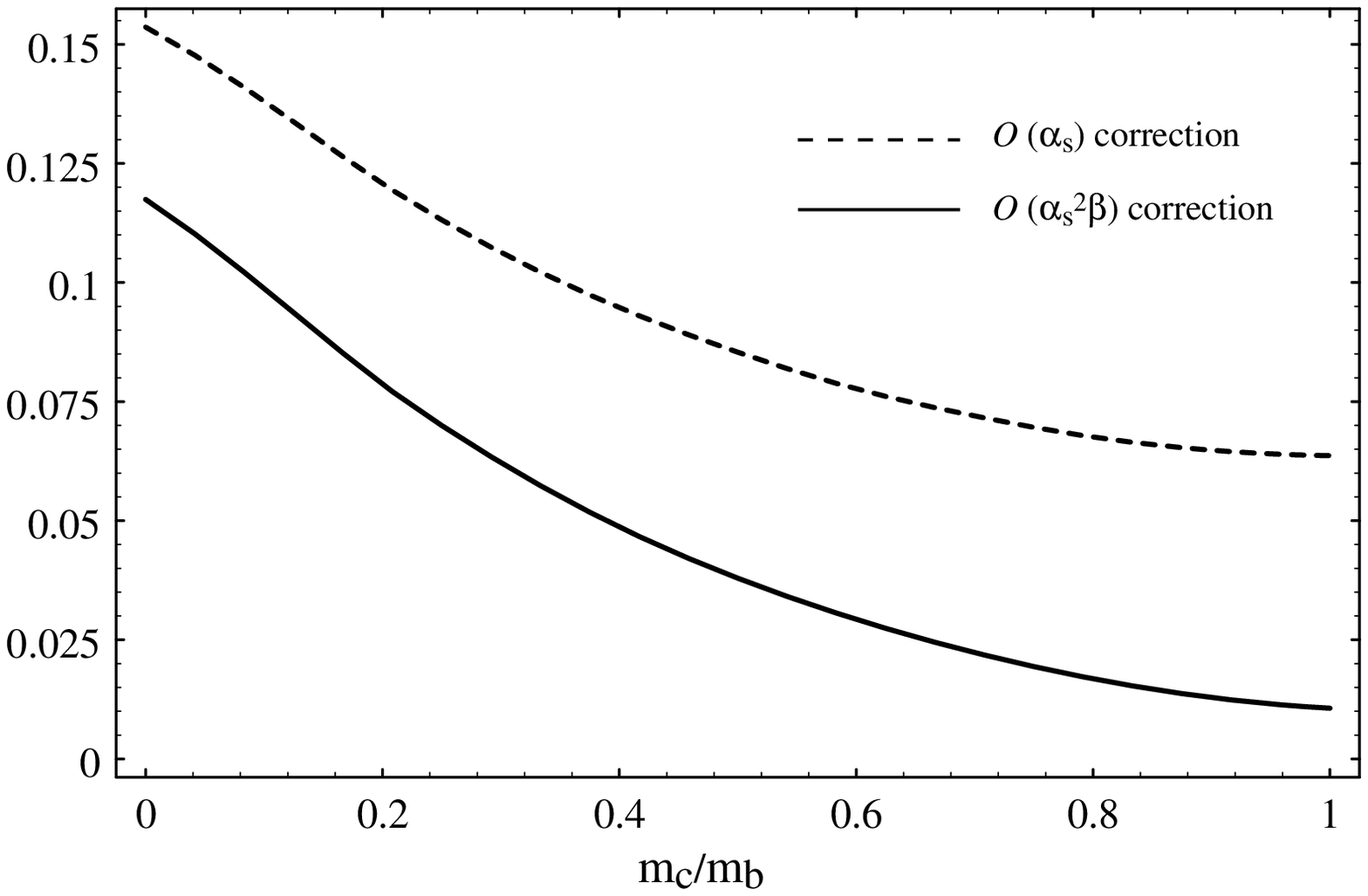}
\insertfig{Figure 4}{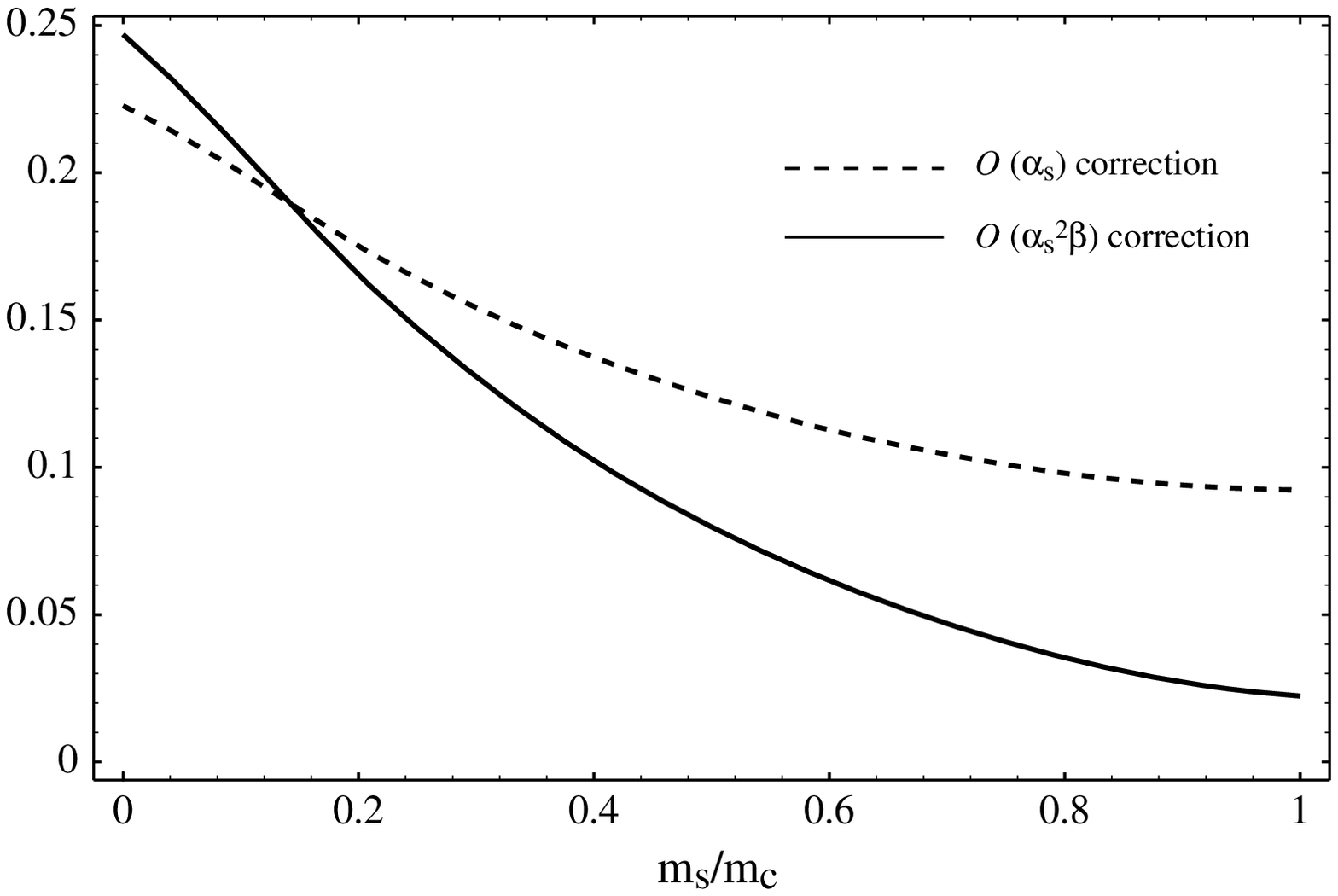}
\vfill\eject
\bye